\documentclass{ws-p8-50x6-00}
\usepackage{amssymb}

\begin{document}

\title{Cosmology of Randall-Sundrum models%
\footnote{Talk given at
``International Workshop on Particle Physics and the Early Universe'',
COSMO2000, Cheju Island, Korea, September 2000}
}
\author{Hang Bae Kim}
\address{
Department of Physics, Lancaster University,
Lancaster LA1 4YB, Great Britain
%\\E-mail: h.kim@lancaster.ac.uk
}
\maketitle
\abstracts{
There are many interesting issues in
the brane world with a large/warped extra dimension.
We focus on the cosmological aspects.
We review the cosmological solutions of the brane world
and how the conventional four-dimensional cosmology is recovered
by including the effect of stabilization.
Its implications on the mass hierarchy and the cosmological constant
are discussed.
}

\section{Introduction}

For past two years, many particle physicists and cosmologists
were excited by the development of two ideas,
the brane world and the warped extra dimensions,
both of which are based on the existence of extra dimensions.
The basic idea of the brane world is that standard model particles are
localized on a (3+1)-dimensional brane (or a set of branes) embedded in
a higher dimensional spacetime, while gravity propagates in the whole
bulk~\cite{brane-world,ADD}.
The warped extra dimension assumes that the background metric is curved
along the extra dimensions, mainly due to the negative bulk
cosmological constant~\cite{Gogberashvili,RS}.
Why are these two ideas so exciting?
They have brought us fresh views and perspectives in
gravity, cosmology, particle physics and string theory.
We have seen many interesting issues discussed so far, such as
the localization of gravity,
the gauge hierarchy problem,
the cosmological constant problem and self-tuning models~\cite{self-tuning},
the construction of supersymmetric RS models and the role of
supersymmetry~\cite{susy-RS},
the connection to string theory or Horava-Witten model
and warped compactification,
the interpretation in light of AdS/CFT holographic duality~\cite{AdSCFT},
the collider signatures of the KK modes and the radion, etc~\cite{collider}.

In this talk, we focus on the cosmological aspect of the two ideas,
mostly in five dimensional models with one extra dimension.
There has been much interest in this because the five dimensional
nature of gravity and the brane setup might lead to the non-conventional
cosmology even at low temperature as well as at high temperature
above TeV scale.
It was found that the Friedmann equation of the brane shows a
$H\propto\rho$ behavior and has an additional dark radiation term~\cite{BDL}.
There was also a difficulty concerning the negative tension brane,
and it was not very clear what happens at temperatures above TeV,
which can alter the early universe cosmology including inflation.

However, in the early models,
a few important ingredients are not addressed, such as
the mechanism for the localization of fields on the brane,
the stabilization of the brane systems~\cite{stabilization} and
the way to achieve necessary fine tunings of parameters.
They inevitably involve bulk matter and dynamics,
and can change the whole picture, for example,
changing the exponential warp factor to power law.
For the cosmological consequences of brane world models,
taking the stabilization into account is found to be
crucial~\cite{KKOP,CGRT,Kim}.
The inclusion of the effect of stabilization recovers
the conventional FRW cosmology at temperatures below TeV.

The talk is organized as follows.
First, we derive the effective four-dimensional brane equations
to see the generalities of brane dynamics,
though its usefulness is limited
by the lack of the knowledge of bulk effects.
Then, we try to solve the five dimensional equations
with an appropriate ansatz.
The solution can be obtained in very restricted cases,
but a framework can be found where we can study brane cosmology
with the effect of stabilization taken into account.
This is done through the $\hat T^5_5$ component
which is adjusted to stabilize the extra dimension
in the presence of brane matter.
Based on this, we analyze the background spacetime
where we also discuss the mass scales and the hierarchy problem,
and one and two brane models in turn.

\section{Effective four-dimensional equations on the brane}

\subsection{Framework}

We consider the five-dimensional spacetime with coordinates $(\tau,x^i,y)$,
and 3-branes embedded in it.
The action describing our framework is
\begin{eqnarray}
S &=& \int_{M}d^5x\sqrt{-\hat g}\left[\frac{M^3}{2}{\hat R}
    -\Lambda_b+{\cal L}_{bM}\right]
\nonumber\\ && +
\sum_{i}\int d^4x
    \sqrt{-g^{(i)}}\left[-\Lambda_i+{\cal L}_{iM}\right]
\label{Action}
\end{eqnarray}
where
$M$ is the fundamental gravitational scale of the model,
$\Lambda_b$ and $\Lambda_i$ represent the bulk cosmological constant and
brane tensions, respectively.
${\cal L}_{bM}$ and ${\cal L}_{iM}$ are Lagrangians for the bulk fields
and for the fields localized in the branes.

To investigate the role of bulk and bulk fields in brane dynamics~\cite{Maeda}
we introduce a bulk scalar $\hat\Phi$,
with ${\cal L}_{bM}=\frac12(\hat\partial\hat\Phi)^2$.
We also allow that $\Lambda_b$, $\Lambda_i$ and ${\cal L}_i$
can be functions of $\hat\Phi$.
The bulk Einstein equations obtained from the action (\ref{Action}) are
\begin{equation}
\label{BEE}
\hat G_{MN} = \hat T_{MN}
= \hat\partial_M\hat\Phi\hat\partial_N\hat\Phi
-\hat g_{MN}\left[\frac12(\hat\partial\hat\Phi)^2+\Lambda_b(\hat\Phi)\right],
\end{equation}
and the bulk scalar equation is
\begin{equation}
\label{BSE}
\hat\square\hat\Phi-\frac{d\Lambda_b(\hat\Phi)}{d\hat\Phi}=0.
\end{equation}

The existence of branes imposes the junction conditions
on the above bulk equation,
giving discontinuities in the first derivatives across the branes
of metric and scalar field. First let us proceed the general setup.
A brane can be described by the normal vector $n_M$.
Then the induced metric on the brane is given by $g_{MN}=\hat g_{MN}-n_Mn_N$,
while the extrinsic curvature by $K_{MN}=\partial_Mn_N$.
The junction conditions are
\begin{eqnarray}
\label{MJC}
[K_{\mu\nu}] &=&
    -\frac{1}{2M^3}\left(T_{\mu\nu}-\frac13g_{\mu\nu}T\right), \\[0mm]
\label{SJC}
[n_M\partial^M\hat\Phi] &=& \frac{1}{2M^3}\left(
    \frac{d\Lambda_i}{d\hat\Phi}
    +\frac{\delta{\cal L}_{iM}}{\delta\hat\Phi} \right),
\end{eqnarray}
where the bracket in the left hand side means the difference across
the brane and $T_{\mu\nu}$ is the energy momentum tensor of brane matter,
{\it i.e.}, $\Lambda_i$ and ${\cal L}_{iM}$.

Suppose that we are localized on a certain brane and
want to study the brane dynamics as observed by us.
We may take two different approaches.
The first approach is to derive the effective 3-brane equations
localized on our brane.
The second is to directly solve the whole bulk equations.
In this section, we follow the first approach.
The second will be dealt with in the next section.

\subsection{The effective 3-brane Einstein equation}

To derive the effective 3-brane equations, we need to know the extrinsic
curvature and the intrinsic curvature of our brane in terms of the bulk
metric $\hat g_{MN}$ and the normal vector $n_{M}$,
which are provided by the Codacci equation
\begin{equation}
\label{Codacci}
\partial_MK^M_\mu-\partial_\mu K = g^M_\mu G_{MN}n^N,
\end{equation}
and the Gauss equation
\begin{eqnarray}
G_{\mu\nu} &=& \frac23 \left[
G_{MN}g^M_\mu g^N_\nu+\left(G_{MN}n^Mn^N-\frac14G\right)g_{\mu\nu} \right]
\nonumber\\ && \label{Gauss}
+KK_{\mu\nu}-K^M_\mu K_{M\nu}-\frac12\left(K^2-K_{MN}K^{MN}\right)g_{\mu\nu}
-E_{\mu\nu},
\end{eqnarray}
where $E_{\mu\nu}=C_{MNOP}n^Mn^Og^N_\mu g^P_\nu$
and $C_{MNOP}$ is the bulk Weyl tensor.

Now we take $y$ as the Gaussian normal coordinates
and impose $Z_2$ symmetry, $y\sim-y$.
We expand $\hat\Phi$ around the brane,
\begin{equation}
\hat\Phi(x,y) = \phi(x)+\Phi_1(x)|y|+\frac12\Phi_2(x)y^2+{\cal O}(y^3).
\end{equation}
Then from the bulk equations (\ref{BEE}) and (\ref{BSE}),
and the junction conditions (\ref{MJC}) and (\ref{SJC}),
we obtain the equation for $\phi$
\begin{equation}
\label{4DESE}
\square\phi-\frac{d\Lambda_b}{d\phi}=
-\frac{T}{12M^3}\left(\frac{d\Lambda}{d\phi}\right)
-\Phi_2,
\end{equation}
and the four-dimensional effective Einstein equation for the brane~\cite{Maeda}
\begin{equation}
\label{4DEEE}
G_{\mu\nu} =
\Lambda_{\rm eff}(\phi)g_{\mu\nu}
+\frac{\Lambda(\phi)}{6M^3}T_{\mu\nu}
+\frac{1}{M^6}\Pi_{\mu\nu}
-E_{\mu\nu}+\frac{2}{3M^3}\hat T_{\mu\nu}(\phi),
\end{equation}
where
\begin{eqnarray}
\Lambda_{\rm eff}(\phi) &=& \frac{1}{2M^3}\left[
\Lambda_b+\frac{\Lambda^2}{6M^3}-\frac18\left(\frac{d\Lambda}{d\phi}\right)^2
\right],
\\
\Pi_{\mu\nu} &=& -\frac14T_{\mu\lambda}T^\lambda_\nu
+\frac{1}{12}TT_{\mu\nu} + \frac18g_{\mu\nu}T_{\lambda\rho}T^{\lambda\rho}
-\frac{1}{24}g_{\mu\nu}T^2,
\\
\hat T_{\mu\nu}(\phi) &=& \partial_\mu\phi\partial_\nu\phi
-\frac58g_{\mu\nu}(\partial\phi)^2.
\end{eqnarray}
The first three terms in the right hand side of (\ref{4DEEE})
deal with the sources on the brane.
The first two terms are same as four-dimensional Einstein gravity,
if we identify the four-dimensional Planck mass
\begin{equation}
\label{4DPM}
M_P^{-2}=\frac{\Lambda(\phi)}{6M^6}.
\end{equation}
The third term gives a correction quadratic in energy-momentum tensor.
The last two terms in Eq.~(\ref{4DEEE}) are bulk effect terms,
which reflect the existence of bulk in brane dynamics.
They are inputs from bulk dynamics and not determined in the brane equations.
The brane equations (\ref{4DESE}) and (\ref{4DEEE}) are not closed equations.
In this sense, they would not be very useful without the knowledge of
the bulk effect terms.
However, they reveal a most general structure of the brane equations.
In this regard, we note that the Planck mass (\ref{4DPM}) seems at first
to be determined solely by the brane tension.
This seems strange because the graviton (and its zero mode)
comes from the bulk fields. Therefore there must be some
correlation between the brane tension and bulk dynamics,
which are incorporated in the brane equations through these bulk effect terms.
We will see this in the following sections where the bulk solutions
are treated.

% \subsection{Applications}
% 
% Minkowski brane world
% \begin{equation}
% \Lambda_{\rm eff}(\phi)=\frac{1}{2M^3}\left[ \Lambda_b+\frac{\Lambda^2}{6M^3}
% -\frac18\left(\frac{d\Lambda}{d\phi}\right)^2 \right] = 0
% \end{equation}
% When $\Lambda_b=0$ (for example, by supersymmetry)
% \begin{equation}
% \frac{\Lambda^2}{6M^3} = \frac18\left(\frac{d\Lambda}{d\phi}\right)^2
% \quad\Rightarrow\quad
% \Lambda\propto e^{\pm2\phi/\sqrt{3M^3}}
% \end{equation}
% Self-tuning solution to the cosmological constant problems\\
% The bulk effect terms will spoil this `good' property.
% 
% FRW brane world
% 
% Limitation of 4D brane equation approach: Bulk Effects!

\section{Five-dimensional cosmological solutions}

\subsection{Framework}

In this section, we investigate the cosmological bulk solutions
of five-dimensional models.
For the two brane models,
the fifth dimension $y$ is assumed to be an orbifold $S^1/Z_2$
with $y\sim y+1$ and $y\sim-y$ identified.
Two 3-branes reside at two fixed points (boundaries) $y=0,\frac12$.

Since we are interested in the cosmological solution,
we consider the metric where the 3-dimensional spatial section is
homogeneous and isotropic.
\begin{equation}
\label{CMA}
ds^2 = -n^2(\tau,y)d\tau^2 + a^2(\tau,y)\gamma_{ij}dx^idx^j + b^2(\tau,y)dy^2,
\end{equation}
where
$\gamma_{ij}$ is the 3-dimensional homogeneous and isotropic metric, and
we will use $K=-1,0,+1$ to represent its spatial curvature.
Einstein equations are given by
$\hat G_{MN}=(1/M^3)\hat T_{MN}$ where
\begin{eqnarray}
\hat G_{00} &=& 3\left\{
 \frac{\dot a}{a}\left(\frac{\dot a}{a}+\frac{\dot b}{b}\right)
-\frac{n^2}{b^2}\left[\frac{a''}{a}
    +\frac{a'}{a}\left(\frac{a'}{a}-\frac{b'}{b}\right)\right]
+ K\frac{n^2}{a^2}\right\} \\
\hat G_{ij} &=& -\frac{a^2}{n^2}\gamma_{ij} \left\{
 2\frac{\ddot a}{a}
+\frac{\dot a}{a}\left(\frac{\dot a}{a}-2\frac{\dot n}{n}\right)
+\frac{\ddot b}{b}
+\frac{\dot b}{b}\left(2\frac{\dot a}{a}-\frac{\dot n}{n}\right)\right\}
-K\gamma_{ij} \nonumber\\
&& +\frac{a^2}{b^2}\gamma_{ij} \left\{
 2\frac{a''}{a}+\frac{n''}{n}
+\frac{a'}{a}\left(\frac{a'}{a}+2\frac{n'}{n}\right)
-\frac{b'}{b}\left(2\frac{a'}{a}+\frac{n'}{n}\right)\right\} \\
\hat G_{55} &=& 3\left\{
-\frac{b^2}{n^2} \left[
\frac{\dot a}{a}\left(\frac{\dot a}{a}-\frac{\dot n}{n}\right)
+\frac{\ddot a}{a}\right] +
\frac{a'}{a}\left(\frac{a'}{a}+\frac{n'}{n}\right) \right\} \\
\hat G_{05} &=& 3\left(\frac{n'}{n}\frac{\dot a}{a} + \frac{a'}{a}\frac{\dot
b}{b} - \frac{\dot a'}{a}\right) \\
\label{EMtensor}
\hat T^M_N &=& {\rm diag}[-\hat\rho,\hat p,\hat p,\hat p,\hat p_5]
+ \sum_{i=0,\frac12}\frac{\delta(y_i)}{b}{\rm diag}[-\rho_i,p_i,p_i,p_i,0]
\end{eqnarray}

In addition to Einstein equations, we use
the energy-momentum conservation equation, $\partial_M\hat T^M_N=0$
\begin{eqnarray}
&\displaystyle
\label{EMC1}
\frac{d\hat\rho}{d\tau} + 3(\hat\rho+\hat p)\frac{\dot a}{a}
+(\hat\rho+\hat p_5)\frac{\dot b}{b} = 0,
&\\&\displaystyle
\label{EMC2}
\hat p_5'+\hat p_5\left(\frac{n'}{n}+3\frac{a'}{a}\right)
+\frac{n'}{n}\hat\rho - 3\frac{a'}{a}\hat\rho = 0.
&
\end{eqnarray}

Brane sources in Eq.~(\ref{EMtensor}) can be converted to
boundary conditions (Junction conditions):
$n$, $a$ and $b$ must be continuous and
$n'$, $a'$ are discontinuous due to boundary sources by
%\footnote{Comment on the higher derivative terms, $R^2$ ...}
\begin{equation}
\label{JC}
\left.\frac{1}{b}\frac{n'}{n}\right|_{y_i-}^{y_i+}=\frac{2\rho_i+3p_i}{3M^3},
\quad
\left.\frac{1}{b}\frac{a'}{a}\right|_{y_i-}^{y_i+}=-\frac{\rho_i}{3M^3}.
\end{equation}

\subsection{Five-dimensional spacetime with the bulk cosmological constant
and brane tensions}

First, we consider the five-dimensional spacetime
supported by the negative bulk cosmological constant
$\hat\rho=-\hat p=\Lambda_b<0$
and brane tensions $\rho_i=-p_i=\Lambda_i$.
We define the parameters
\begin{equation}
k=\left(\frac{-\Lambda_b}{6M^3}\right)^{1/2},\quad
k_i=\frac{\Lambda_i}{6M^3}
\end{equation}
In general cases with $k_0,-k_{1/2}\ge k$,
the metric is given by~\cite{KK}
\begin{equation}
\label{KK}
ds^2 = \frac{-d\tau^2+\delta_{ij}dx^idx^j+(k\tau b_0)^2dy^2}%
{\left[k\tau\sinh(kb_0|y|+c_0)+g_0\right]^2}
\end{equation}
where $b_0$ and $c_0$ are determined by
\begin{equation}
k\cosh(c_0) = k_0 ,\qquad
k\cosh(\frac12kb_0+c_0) = -k_{\frac12}.
\end{equation}
The metric (\ref{KK}) describes a slice of AdS$_5$ space
inflating both in extra dimension and in spatial dimensions.
We note two special cases.
If we have a fine tuning
$(-k_2-k)/(k_1-k)=e^{kb_o}(-k_2+k)/(k_1+k)$,
we obtain a inflationary solution with static extra dimension~\cite{Kaloper}
\begin{equation}
ds^2 = \frac{-d\tau^2+e^{2k\tau}\delta_{ij}dx^idx^j+b_0^2dy^2}%
{\sinh^2(kb_0|y|+c_0)}
\end{equation}
With two fine tunings $k=k_0=-k_{\frac12}$,
we can get a static solution, the Randall-Sundrum model~\cite{RS}
\begin{equation}
\label{RS}
ds^2 = e^{-2kb_0|y|}\eta_{\mu\nu}dx^\mu dx^\nu + b_0^2dy^2
\end{equation}
This model attracted much attention because it convert the gauge hierarchy
problem into a geometric problem.
The four-dimensional Planck scale in this model is given by
$M_P^2=(M^3/k)[1-e^{-kb_0}]$.
Any mass parameter $m_0$ on the visible brane corresponds to a physical mass
$m=m_0e^{-\frac12kb_0}$, which we identify as the weak scale.
Hence the large hierarchy between the Planck scale and the weak scale
$M_P/M_W\sim10^{16}$ can be explained by a warp factor $e^{-\frac12kb_0}$
with $\frac12kb_0\sim37$.
The property is not spoiled by $R^2$ corrections,
if $R^2$ corrections are given by Gauss-Bonnet interactions~\cite{KKL}.

Note that the space described by the metric (\ref{KK}) is locally AdS$_5$,
because it shares the same bulk equation.
The metric (\ref{KK}) can be transformed to (\ref{RS})
by a coordinate transformation
\begin{equation}
e^{-kb_{\rm RS}y_{\rm RS}}=k\tau\sinh(kb_0y+c_0)+g_0,\quad
\tau_{\rm RS}=\tau\cosh(kb_0y+c_0).
\end{equation}
Therefore, we can say that above metrics are the slices of AdS$_5$
with different boundary geometries.

\subsection{Cosmological solutions with static extra dimension}

Let us turn to the more interesting situation
where the matter is added on the brane(s) and possibly in the bulk.
But we require the extra dimension to be static, that is $\dot b=0$.
This requires a fine tuning between matter densities.
For the bulk energy-momentum, we assume
\begin{equation}
\hat\rho=\Lambda_b,\quad
\hat p=-\Lambda_b,\quad
\hat p_5=-\Lambda_b+p_5(\tau,y).
\end{equation}
where $\Lambda_b<0$.
The form of $p_5(\tau,y)$ is constrained by Eq.~(\ref{EMC2}) to be
\begin{equation}
\displaystyle
p_5(\tau,y) = \frac{\tilde p_5(\tau)}{n(\tau,y)a^3(\tau,y)}
\end{equation}
With the gauge fixing $n(\tau,y=0)=1$,
we obtain the solution~\cite{BDEL,Kim}
\begin{eqnarray}
a^2(\tau,y) &=& a_0^2\left[
\left(1+\frac{\dot a_0^2+K}{2k^2a_0^2}\right)\cosh(2kby)
-\frac{\dot a_0^2+K}{2k^2a_0^2}
\right.\nonumber\\[5mm] && \hspace{10mm} \left.
\pm \left(1+\frac{1}{k^2}
\left[\frac{\dot a_0^2+K}{a_0^2}+\frac{C}{a_0^4}\right]
\right)^{1/2}\sinh(2kby) \right],
\end{eqnarray}
and $n(\tau,y)=\dot a(\tau,y)/\dot a_0(\tau)$.
Here $a_0(\tau)\equiv a(\tau,y=0)$ and 
$C(\tau)$ is determined by $\tilde p_5(\tau)$
up to a constant through the equation
\begin{equation}
\label{CE}
\dot C = \frac{2\dot a_0(\tau)}{3M^3}\ \tilde p_5(\tau).
\end{equation}
$a_0(\tau)$ is fixed by the boundary condition.
Suppose that a brane with the energy density $\rho_0$ is placed at $y=0$,
and assume $Z_2$ symmetry $y\sim-y$.
The junction condition (\ref{JC}) at the brane gives the
evolution equation for $a_0(\tau)$
\begin{equation}
\label{SFEE}
\left(\frac{\dot a_0}{a_0}\right)^2+\frac{K}{a_0^2}
= -k^2 + \left(\frac{\rho_0}{6M^3}\right)^2 - \frac{C}{a_0^4}
\end{equation}
If we further assume that we are living on the brane,
this equation is nothing but the Friedmann equation of our universe.
However, in the simple case where $\Lambda_b=0$ and $p_5=0$,
this equation differs from the four-dimensional Friedmann equation
in two aspects. First, the Hubble parameter $H=\dot a_0/a_0$
is proportional to $\rho_0$ instead of $\sqrt{\rho_0}$.
Second, there is a $C/a_0^4$ term which looks like a radiation term.
This means that we can have dynamic solution without any matter
on the brane or in the bulk.
This term seems to have an interesting interpretation in view of
AdS/CFT~\cite{AdSCFT}.
Both of these alter the big bang nucleosynthesis result:
$H\propto\rho_0$ is not compatible and $C$ is required to be small enough.

If we consider the negative bulk cosmological constant
together with the positive brane tension, $\rho_0=\Lambda_0+\rho_{0M}$,
the above equation is written as
\begin{equation}
\left(\frac{\dot a_0}{a_0}\right)^2+\frac{K}{a_0^2}
= (k_0^2-k^2) %+ \left(\frac{\Lambda_0}{6M^3}\right)^2
+ \frac{\Lambda_0}{18M^6}\rho_{0M}
+ \frac{1}{36M^6}\rho_{0M}^2
- \frac{C}{a_0^4}.
\end{equation}
The four-dimensional Friedmann equation is obtained
for $k_0^2-k^2=0$, $C=0$, $\rho_{0M}\ll\Lambda_0$.
We can obtain a viable cosmology for the positive tension brane attached to
the infinite size extra dimension.
The effective cosmological constant is given by
$\Lambda_{\rm eff}\propto(k_0^2-k^2)$
At high energy/temperature, that is, in the very early universe
$\rho_{0M}^2$ term dominates and results in very interesting
consequences~\cite{MWBH}.

%For the negative tension brane, the metric ansatz (\ref{CMA})
%is not valid.

\section{The stabilized RS models}

\subsection{Stabilization by balanced bulk matter}

In the previous section, we saw that the brane at $y=0$ fixes
the whole bulk solution.
If we have the second brane at $y=\frac12$,
the junction condition at $y=\frac12$ gives
\begin{equation}
\label{EDC}
\bar\rho_{\frac12} = \frac
{\sinh(kb)+\frac12(\bar\rho_0^2-\bar C-1)\sinh(kb)-\bar\rho_0\cosh(kb)}
{\cosh(kb)+\frac12(\bar\rho_0^2-\bar C-1)(\cosh(kb)-1)-\bar\rho_0\sinh(kb)},
\end{equation}
where
\begin{equation}
\bar\rho_i=\frac{\rho_i}{6M^3k}\ ({\textstyle i=0,\frac12}),\quad
\bar C=\frac{C}{k^2a_0^4}.
\end{equation}
This is a constraint between energy densities $\rho_i$ on two branes.
The reason why this constraint is necessary is obvious.
We used the ansatz with the static extra dimension,
which is not the general case for the two brane model.
But the (almost) static extra dimension is required
from the phenomenological view point.
We need a stabilization mechanism to make the extra dimension static.

Here we consider a simple modeling of how the stabilization mechanism
works~\cite{Kim,KKOP}.
It must involve some bulk dynamics.
We suppose that it is through the role of $C$.
The basic idea is that the back reaction by the stabilization mechanism
to the brane energy densities which, if left alone,
would destabilize the extra dimension,
induces the bulk energy momentum $\hat T^5_5$
which forces $C$ fitted to the constraint (\ref{EDC}) through (\ref{CE})
and keep $\dot b=0$.
The constraint (\ref{EDC}) can be solved to give the necessary $C$
\begin{equation}
\label{FC}
\bar C = (\bar\rho_0^2-1) - 2\,
\frac{\bar\rho_0+\bar\rho_{\frac12}-(1+\bar\rho_0\bar\rho_{\frac12})\tanh(kb)}
{\tanh(kb)\{1-\bar\rho_{\frac12}\tanh(kb/2)\}},
\end{equation}
and the bulk energy-momentum component $p_5$ through
\begin{equation}
\label{P5CE}
p_5(\tau,y) = \frac{6M^3}{4a^3(\tau,y)\dot a(\tau,y)}\;\dot C.
\end{equation}
Inserting (\ref{FC}) into (\ref{SFEE}), we obtain
the 3-brane Friedmann equation for the stabilized two brane system
\begin{equation}
\label{3BFE}
\left(\frac{\dot a_0}{a_0}\right)^2+\frac{K}{a_0^2} = 2k^2\,
\frac{\bar\rho_0+\bar\rho_{\frac12}-(1+\bar\rho_0\bar\rho_{\frac12})\tanh(kb)}
{\tanh(kb)\{1-\bar\rho_{\frac12}\tanh(kb/2)\}}.
\end{equation}

We have two comments here.
First, in general it is expected that the stabilization mechanism
induces $\hat T^i_i$ as well as $\hat T^5_5$,
and $b$ may not be strictly static but be shifted somewhat
as $\rho_i$ changes in time.
Considering $\hat T^5_5$ only and requiring strictly static $b$
seems to give a limit of infinitely steep stabilization potential.
It is of course unrealistic, but gives the leading behaviors
of RS models with a certain, unknown stabilization mechanism.
Second, we can see from (\ref{P5CE}) that
the induced $p_5$ does not fix the additive constant of $C$.
This cannot be controlled by $\hat T^5_5$, but required to vanish
to satisfy the constraint. This may require another mechanism behind.
Actually non-vanishing $C$ implies the breakdown of conformal symmetry
in the bulk, and there might be a connection between the stabilization
mechanism which requires non-trivial $C$ and
the conformal symmetry breaking.

Now we rephrase the metric for the stabilized RS model
\begin{equation}
\label{SRSM1}
ds^2 = -n^2(\tau,y)d\tau^2+a^2(\tau,y)\delta_{ij}dx^idx^j+b^2(\tau,y)dy^2
\end{equation}
where
\begin{eqnarray}
\label{SRSM2}
b(\tau,y) &=& b = \hbox{constant}, \nonumber\\
n(\tau,y) &=& \frac{\dot a(\tau,y)}{\dot a(\tau,0)}, \nonumber\\
a(\tau,y) &=& a_0(\tau)\left[
\cosh(2kby)-\bar\rho_0\sinh(2kby)
\vphantom{\frac{\bar\rho_{\frac12}}{\bar\rho_{\frac12}}}
\right.\nonumber\\ &&\hspace{0mm}\left.
+\frac{(\bar\rho_0+\bar\rho_{\frac12})\cosh(kb)
  -(1+\bar\rho_0\bar\rho_{\frac12})}
{\sinh(kb)-\bar\rho_{\frac12}\{\cosh(kb)-1\}}\{\cosh(2kby)-1\} \right]^{1/2}
\end{eqnarray}
and $a_0(\tau)$ satisfies (\ref{3BFE}).

% The ratio of the scale factor at two branes
% \begin{equation}
% \frac{a^2(\tau,\frac12)}{a^2(\tau,0)} =
% \frac{1-\bar\rho_0\tanh(kb/2)}{1-\bar\rho_{\frac12}\tanh(kb/2)}
% \end{equation}
% The ratio of expansion rates of two branes
% \begin{equation}
% \frac{\dot a(\tau,\frac12)/a(\tau,\frac12)}{\dot a(\tau,0)/a(\tau,0)} =
% \frac{2+(\bar\rho_0+3\bar p_0)\tanh(kb/2)}
% {2+(\bar\rho_{\frac12}+3\bar p_{\frac12})\tanh(kb/2)} \;
% \frac{1-\bar\rho_{\frac12}\tanh(kb/2)}{1-\bar\rho_0\tanh(kb/2)}
% \end{equation}
% The ratio of lapse function $n(\tau,y)$ at two branes
% \begin{equation}
% \frac{n(\tau,\frac12)}{n(\tau,0)} =  
% \frac{2+(\bar\rho_0+3\bar p_0)\tanh(kb/2)}
% {2+(\bar\rho_{\frac12}+3\bar p_{\frac12})\tanh(kb/2)}
% \left[\frac{1-\bar\rho_{\frac12}\tanh(kb/2)}{1-\bar\rho_0\tanh(kb/2)}
% \right]^{1/2}
% \end{equation}

\subsection{The static five-dimensional spacetime:
The mass hierarchy and the cosmological constant}

With the metric in (\ref{SRSM1}) and (\ref{SRSM2}), we first consider
the case where there is no matter and only the bulk cosmological constant
and the brane tensions are involved.
The metric is now given by $a(\tau,y)=a_0(\tau)n(y)$ and
\begin{eqnarray}
\label{N2}
n(y) &=& \left[
\cosh(2kby)-\bar k_0\sinh(2kby)
\vphantom{\frac{\bar k_{\frac12}}{\bar k_{\frac12}}}
\right.\nonumber\\ &&\hspace{0mm}\left.
+\frac{(\bar k_0+\bar k_{\frac12})\cosh(kb)
  -(1+\bar k_0\bar k_{\frac12})}
{\sinh(kb)-\bar k_{\frac12}\{\cosh(kb)-1\}}\{\cosh(2kby)-1\} \right]^{1/2}.
\end{eqnarray}
From (\ref{FC}) and (\ref{P5CE}), the balanced $p_5$ is found to be
\begin{equation}
\label{P52}
p_5(y) = \frac{6M^3k^2}{n(y)^4} \left[ (\bar k_0^2-1) - 2
\frac{(\bar k_0+\bar k_{\frac12})-(1+\bar k_0\bar k_{\frac12})\tanh(kb)}
{\tanh(kb)\{1-\bar k_{\frac12}\tanh(kb/2)\}} \right].
\end{equation}
The scale factor $a(\tau,y)$ undergoes inflation,
and the Hubble parameter can be defined independently of $y$
because $H=\dot a(\tau,y)/a(\tau,y)=\dot a_0(\tau)/a_0(\tau)$.

The static background spacetime is obtained when we make a fine tuning
to satisfy the condition for the vanishing cosmological constant
\begin{equation}
\label{VCCC}
(\bar k_0+\bar k_{\frac12})-(1+\bar k_0\bar k_{\frac12})\tanh(kb) = 0.
\end{equation}
With this condition, (\ref{N2}) and (\ref{P52}) are simplified to
\begin{eqnarray}
n(y) &=& \left[\cosh(2kby)-\bar k_0\sinh(2kby)\right]^{1/2}, \\
\label{P53}
p_5(y) &=& \frac{6M^3(k_0^2-k^2)}{n(y)^4}.
\end{eqnarray}
We can identify the four-dimensional Planck scale in two ways.
Firstly, we can get it from the 4-dimensional effective theory
which is obtained by integrating the action over the extra dimension,
\begin{equation}
\label{MP1}
M_P^2 = M^3 \int_{-\frac12}^{\frac12}b\,dy\,n(y)^2
= \frac{M^3}{k}\left[\sinh(kb)-\bar k_0\{\cosh(kb)-1\}\right].
\end{equation}
Secondly, we can deduce it from the 3-brane Friedmann equation
(\ref{3BFE}), which leads to
\begin{equation}
\label{MP2}
M_P^2 = \frac{M^3}{k} \,
\frac{\tanh(kb)\{1-\bar k_{\frac12}\tanh(kb/2)\}}
{1-\bar k_{\frac12}\tanh(kb)}
\end{equation}
The two derived Planck scales (\ref{MP1}) and (\ref{MP2}) coincide
under the condition (\ref{VCCC}),
that is, when the cosmological constant vanishes.

Let us consider the gauge hierarchy problem in this background spacetime.
Physical mass scale at two branes are given by~\cite{KKOP}
\begin{eqnarray}
M_W &=& Mn(\tau,0) = M, \\
M_H &=& Mn(\tau,\frac12) = M\left[\cosh(kb)-\bar k_0\sinh(kb)\right]^{1/2}.
\end{eqnarray}
Note that we placed the visible brane at $y=0$ and identified
the physical mass scale of the visible brane as the weak scale.
The physical mass scale on the hidden brane, called the hidden scale here,
is in general different from the Planck scale.
Now the ratio of the electroweak scale and the Planck scale is
\begin{equation}
\label{GH}
\frac{M_P^2}{M_W^2} = \left(\frac{M}{k}\right)
\left[\sinh(kb)-\bar k_0\left\{\cosh(kb)-1\right\}\right]
\approx 10^{32}.
\end{equation}
The condition for the vanishing cosmological constant and
the solution to the gauge hierarchy problem impose two conditions
among four parameters $k$, $k_0$, $k_{\frac12}$ and $b$.
Hence, in this model, there are continuous set of static background
spacetimes which solve the cosmological constant problem and
the gauge hierarchy problem together,
specified by, for example, two parameters $(k/M,kb)$
and $k_0$ and $k_{\frac12}$ can be expressed in terms of them
from (\ref{VCCC}) and (\ref{GH})
\begin{eqnarray}
\bar k_0 &=& \frac{\sinh(kb)-10^{32}\left(k/M\right)}{\cosh(kb)-1}, \\
\bar k_{\frac12} &=& \frac{-\bar k_0+\tanh(kb)}{1-\bar k_0\tanh(kb)}.
\end{eqnarray}
The original RS model with $k=-k_0=k_{\frac12}$ is a special case
where $p_5$ in (\ref{P53}) vanishes.

% Comment on the hidden scale:
Note that while the hidden scale is larger than the weak scale
by the ratio of warp factors at two branes, $n(\tau,\frac12)/n(\tau,0)$,
it is in general different from the four-dimensional Planck scale
by a factor $M/k$ and a different warp factor combination.
In the original RS model where $k=-k_0=k_{\frac12}$, $e^{kb}\gg1$ and
$M/k\sim1$, the difference disappears.
If we introduce another hierarchy $M/k\gg1$
(without any feasible reason at the moment),
this will show up in the hierarchy of the hidden scale and the Planck scale.

\subsection{One brane model}

Let us turn to the cases where matters are added on the brane.
First, we look at the case $\rho_{\frac12}=0$.
This corresponds to compactifying the extra dimension without the second
brane. This is made possible by the tuned $\hat p_5$ distribution along
the extra dimension
\begin{equation}
\hat p_5 = \frac{a_0^3}{na^3}\left[
    6M^3k^2 - \frac12 k{\rm coth}(kb)(\rho_0-3p_0)
    - \frac{\rho_0(\rho_0+3p_0)}{12M^3}
\right].
\end{equation}
The 3-brane Friedmann equation (\ref{3BFE}) becomes
\begin{equation}
\left(\frac{\dot a_0}{a_0}\right)^2+\frac{K}{a_0^2} =
2k^2\left[-1+\bar\rho_0{\rm coth}(kb)\right].
\end{equation}
If we split out the brane tension from the brane energy density,
the above equation takes the form of four-dimensional Friedmann equation
\begin{equation}
\left(\frac{\dot a_0}{a_0}\right)^2+\frac{K}{a_0^2} =
2k^2\left[\left(\frac{k_0}{k}{\rm coth}(kb)-1\right) +
{\rm coth}(kb)\bar\rho_{0M}\right],
\end{equation}
with the identifications
\begin{eqnarray}
M_p^2 &=& \frac{M^3}{k}\tanh(kb), \\
\Lambda_{\rm eff} &=& \Lambda_0-(6M^3\Lambda_b)^{\frac12}\tanh(kb).
\end{eqnarray}

An interesting characteristic of this model is its implication
for the cosmological constant problem.
Suppose that we have the relation $k=k_0$ in some way, that is,
assume a solution to the `big' cosmological constant problem.
Then the size of the cosmological constant has an exponential dependence
on the size of extra dimension.
This is just the RS-type solution to the `small' cosmological constant
problem. The currently observed cosmological constant
$\Omega_\Lambda\sim1$ can be fitted with $kb\approx140$.

\subsection{Two brane model}

Now we turn to the two brane case.
We split the brane energy densities into the brane tensions and the brane
matter energy densities as we did in the one brane model,
and impose the vanishing cosmological constant condition (\ref{VCCC}).
The the 3-brane Friedmann equation (\ref{3BFE}) can be written as
\begin{equation}
\left(\frac{\dot a_0}{a_0}\right)^2+\frac{K}{a_0^2} = \frac
{\rho_{0M}+\tilde\rho_{\frac12M}
    \displaystyle
    +\frac{\rho_{0M}\tilde\rho_{\frac12M}}{6M^3k}\,
    \frac{\tanh(kb)}{1-\bar k_0\tanh(kb)}}
{3M_P^2\left[1
    \displaystyle
    -\frac{\tilde\rho_{\frac12M}}{6M_P^2k^2}\,
    \frac{\tanh(kb)\tanh(kb/2)}{1-\bar k_0\tanh(kb)}\right]},
\end{equation}
where $\tilde\rho_{\frac12M}=\rho_{\frac12M}n(\frac12)^4
=\rho_{\frac12M}\left[\cosh(kb)-\bar k_0\sinh(kb)\right]^2$,
which is the physically observed energy density on the hidden brane.
Up to small correction, this equation is nothing but the four-dimensional
Friedmann equation. The energy densities on both branes contribute equally.
So the matter on the hidden brane acts as dark matter for our brane.

The inclusion of the effect of stabilization mechanism,
through the balanced $\hat T^5_5$ component in this simplified model,
gives a ordinary FRW cosmology at least up to TeV scale,
resolving the peculiarities caused by the existence of
the negative tension brane.
Above the TeV scale, we meet a complicated situation
where in addition that the quadratic terms become important,
we need to consider the excitation of other dynamical variables
which may spoil the stabilization.
Further study is required to clarify it.

\section{Conclusion}

There are many interesting issues
in the brane world models and large/warped extra dimensions,
such as
the mass hierarchy,
the cosmological constant,
localization of gravity,
confinement of fields on the brane,
%(Horava-Witten, D-branes, Topological defects),
fine tunings of the bulk cosmological constant and brane tensions,
%(Supersymmetry, Holography),
stabilization of the extra dimension,
%(Goldberger-Wise),
the role of supersymmetry,
%(Supersymmetric RS),
the role of AdS/CFT duality,
connection to string theory,
%(Horava-Witten, Warped compactifications),
and collider signals, etc.
%(KK modes, Radion).

In this talk, we focused on the cosmological implications,
together with the cosmological constant and the gauge hierarchy.
In cosmological side,
the model with one positive tension brane and the infinite warped
extra dimension (RS2) has a viable cosmology
without the need of stabilization.
But for models with two or more branes or with the compact extra dimension,
taking the effects of stabilization into account is very crucial
in studying the cosmology of the models.
We have shown that, through the simple method using the balaced
$\hat T^5_5$ component, the inclusion of the effects of stabilization
recovers the conventional four-dimensional FRW cosmology.
Therefore, the stabilized RS models have viable cosmology below TeV scale,
with interesting new perspective on the mass hierarchy and the
cosmological constant.
Further study is required to clarify the cosmology of these models
at temperatures above TeV scale.

\section*{Acknowledgments}

I would like to express my gratitute to the organizers of the
``International Workshop on Particle Physics and the Early Universe'',
COSMO2000 and acknowledge the hospitality of KIAS.
I have benefited from discussions with K. Choi and H. D. Kim.


\begin{thebibliography}{99}

\newcommand{\journal}[6]{#1 {\bf #2}, #3 (#4)}
\newcommand{\ibid}[5]{{\it ibid.\/} {\bf #1}, #2 (#3)}
\newcommand{\arxiv}[2]{{\tt#1}}

\newcommand{\apj}[5]{Astrophys.\ J.\ {\bf #1}, #2 (#3)}
\newcommand{\epl}[5]{Europhys.\ Lett.\ {\bf #1}, #2 (#3)}
\newcommand{\jhep}[5]{JHEP {\bf #1}, #2 (#3)}
\newcommand{\lnp}[5]{Lect.\ Notes Phys.\ {\bf #1}, #2 (#3)}
\newcommand{\mpl}[5]{Mod.\ Phys.\ Lett.\ {\bf #1}, #2 (#3)}
\newcommand{\nat}[5]{Nature\ {\bf #1}, #2 (#3)}
\newcommand{\npb}[5]{Nucl.\ Phys.\ {\bf B#1}, #2 (#3)}
\newcommand{\plb}[5]{Phys.\ Lett.\ {\bf B#1}, #2 (#3)}
\newcommand{\prd}[5]{Phys.\ Rev.\ {\bf D#1}, #2 (#3)}
\newcommand{\prl}[5]{Phys.\ Rev.\ Lett.\ {\bf #1}, #2 (#3)}
\newcommand{\prt}[5]{Phys.\ Rep.\ {\bf #1}, #2 (#3)}
\newcommand{\zp}[5]{Z.\ Phys.\ {\bf #1}, #2 (#3)}

%% The orginal idea of brane world

\bibitem{brane-world}
V. A. Rubakov and M. E. Shaposhnikov,
\plb{125}{136}{1983}%
{Do we live inside a domain wall?};
\plb{125}{139}{1983}%
{Extra space-time dimensions: Towards a solution to the cosmological constant
problem};
K. Akama, \lnp{176}{267}{1982}{hep-th/0001113}%
{Pregeometry}.

%% Large extra dimensions

\bibitem{ADD}
N. Arkani-Hamed, S. Dimopoulos and G. Dvali,
\plb{429}{263}{1998}{hep-ph/9803315}%
{The hierarchy problem and new dimensions at a millimeter};
I. Antoniadis, N. Arkani-Hamed, S. Dimopoulos and G. Dvali,
\ibid{436}{257}{1998}{hep-ph/9804398}%
{New dimensions at a millimeter to a Fermi and superstrings at a TeV}.

%% Warped extra dimension

\bibitem{Gogberashvili}
M. Gogberashvili,
\arxiv{hep-ph/9812296}%
{Hierarchy problem in the shell-Universe model};
\epl{49}{396}{2000}{hep-ph/9812365}%
{Our world as an expanding shell};
\mpl{A14}{2025}{1999}{hep-ph/9904383}%
{Four Dimensionality in Non-compact Kaluza-Klein Model}.

\bibitem{RS}
L. Randall and R. Sundrum,
\prl{83}{3370}{1999}{hep-ph/9905221}%
{A large mass hierarchy from a small extra dimension};
\prl{83}{4690}{1999}{hep-th/9906064}%
{An alternative to compactification}

%% Self-tuning models

\bibitem{self-tuning}
N. Arkani-Hamed, S. Dimopoulos, N. Kaloper and R. Sundrum,
\plb{480}{193}{2000}{hep-th/0001197}%
{A small cosmological constant from a large extra dimension};
S. Kachru, M. Schulz and E. Silverstein,
\prd{62}{045021}{2000}{hep-th/0001206}%
{Self-tuning flat domain walls in 5d gravity and string theory};
S. F\"orste, Z. Lalak, S. Lavignac and H. P. Nilles,
\plb{481}{360}{2000}{hep-th/0002164}%
{A comment on self-tuning and vanishing cosmological constant
in the brane world};
\jhep{0009}{034}{2000}{hep-th/0006139}%
{The cosmological constant problem from a brane-world perspective};
J. E. Kim, B. Kyae, H. M. Lee,
\arxiv{hep-th/0011118}%
{A model for selftuning the cosmological constant};
\arxiv{hep-th/0101027}%
{Selftuning solution of the cosmological constant problem with antisymmetric
tensor field}.

%% Supersymmetric RS

\bibitem{susy-RS}
T. Gherghetta and A. Pomarol,
\npb{586}{141}{2000}{hep-ph/0003129}%
{Bulk fields and supersymmetry in a slice of ads};
A. Falkowski, Z. Lalak and S. Pokorski,
\plb{491}{172}{2000}{hep-th/0004093}%
{Supersymmetrizing branes with bulk in five-dimensional supergravity};
E. Bergshoeff, R. Kallosh and A. Van Proeyen,
\jhep{0010}{033}{2000}{hep-th/0007044}%
{Supersymmetry in singular spaces};
M. Zucker,
\arxiv{hep-th/0009083}%
{Supersymmetric brane world scenarios From off-Shell supergravity}.

%% AdS/CFT, Holography

\bibitem{AdSCFT}
%P. Kraus, \jhep{9912}{011}{1999}{hep-th/9910149}%
%{Dynamics of anti-de Sitter domain walls};
S. S. Gubser, \arxiv{hep-th/9912001}%
{AdS/CFT and gravity};
N. Arkani-Hamed, M. Porrati, and L. Randall,
\arxiv{hep-th/0012148}%
{Holography and Phenomenology};
R. Rattazzi and A. Zaffaroni, \arxiv{hep-th/0012248}%
{Comments on the holographic picture of the Randall-Sundrum model}.

%% Collider signals

\bibitem{collider}
C. Cs\'aki, M. L. Graesser and G. D. Kribs,
\prd{63}{065002}{2001}{hep-th/0008151}%
{Radion dynamics and electroweak physics};
S. C. Park, H. S. Song and J. Song,
\arxiv{hep-ph/0009245}%
{Radion effects on the production
of an intermediate mass scalar and z at LEP-2};

\bibitem{BDL}
P. Bin\'etruy, C. Deffayet, and D. Langlois, \npb{565}{269}{2000}{}%
{Non-conventional cosmology from a brane universe}

%% Bulk scalar, Goldberger-Wise

\bibitem{stabilization}
W.D. Goldberger and M.B. Wise,
\prl{83}{4922}{1999}{hep-ph/9907447}%
{Modulus stabilization with bulk fields}
O. DeWolfe, D. Z. Freedman, S.S. Gubser, and A. Karch,
\prd{62}{046008}{2000}{}%
{Modeling the fifth-dimension with scalars and gravity}

%% RS cosmology with stabilization

\bibitem{KKOP}
P. Kanti, K. A. Olive and M. Pospelov,
\prd{62}{126004}{2000}{hep-ph/0005146}%
{Solving the hierarchy problem in two-brane cosmological models};
P. Kanti, I. I. Kogan, K. A. Olive and M. Pospelov,
\plb{468}{31}{1999}{}%
{Cosmological three-brane solutions};
\prd{61}{106044}{2000}{}%
{Single brane cosmological solutions with a stable compact extra dimension};
\plb{481}{386}{2000}{}%
{Static solutions for brane models with a bulk scalar field}.

\bibitem{CGRT}
C. Csaki, M. Graesser, L. Randall and J. Terning,
\prd{62}{045015}{2000}{hep-ph/9911406}%
{Cosmology of brane models with radion stabilization}
J. M. Cline and H. Firouzjahi,
\plb{495}{271}{2000}{hep-th/0008185}%
{5-dimensional warped cosmological solutions with radius
stabilization by a bulk scalar}

\bibitem{Kim}
H. B. Kim, \plb{478}{285}{2000}{hep-ph/0001209}%
{Cosmology of Randall-Sundrum models with an extra dimension
stabilized by balancing bulk matter}


%% Effective 4-dimensional equations on the brane

\bibitem{Maeda}
K. Maeda and D. Wands,
\prd{62}{124009}{2000}{}%
{Dilaton gravity on the brane};
T. Shiromizu, K. Maeda and M. Sasaki,
\prd{62}{024012}{2000}{}%
{The Einstein equation on the 3-brane world}.

%% Inflationary solutions

\bibitem{KK}
H. B. Kim and H. D. Kim, \prd{61}{064003}{2000}{}%
{Inflation and gauge hierarchy in Randall-Sundrum compactification}.

\bibitem{Kaloper}
N. Kaloper, \prd{60}{123506}{1999}{}%
{Bent domain walls as brane-worlds};
T. Nihei, \plb{465}{81}{1999}{}%
{Inflation in the five-dimensional universe with an orbifold extra dimension}.

%% Effect of Gauss-Bonet term

\bibitem{KKL}
J. E. Kim, B. Kyae and H. M. Lee,
\prd{62}{045013}{2000}{}%
{Effective Gauss-Bonnet interaction in Randall-Sundrum compactification};
\npb{582}{296}{2000}{}%
{Various modified solutions of the Randall-Sundrum model
with the Gauss-Bonnet interaction}

\bibitem{BDEL}
P. Bin\'etruy, C. Deffayet, U. Ellwanger, and D. Langlois,
\plb{477}{285}{2000}{}%
{Brane cosmological evolution in a bulk with cosmological constant}

\bibitem{MWBH}
R. Maartens, D. Wands, B. A. Bassett, I. Heard, \prd{62}{041301}{2000}{}%
{Chaotic inflation on the brane}


\end{thebibliography}
\end{document}